\documentclass[aps,prl,twocolumn,superscriptaddress,longbibliography, raggedbottom, nobalancelastpage]{revtex4-2}

\usepackage{amsmath,amssymb,amsfonts,bm}
\usepackage{graphicx}
\usepackage{xcolor}
\usepackage[normalem]{ulem}
\usepackage{cancel}
\usepackage{braket}
\usepackage{hyperref}
\hypersetup{
	colorlinks=true,
	citecolor=blue,
	urlcolor=blue,
	linkcolor=blue
}

\definecolor{crimson}{RGB}{255,102,255}
\definecolor{gbgreen}{RGB}{0,100,0}

\newcommand{\gb}[1]{{\color{gbgreen}\ifmmode #1\else[\textbf{gb:} #1]\fi}}
\definecolor{fixorange}{RGB}{214,110,0}

\allowdisplaybreaks

\begin{document}
	
	\title{
		Thermo-electric Response Bound on Fluctuations in Coherent Transport
	}

	\author{Gianmichele Blasi}
	\affiliation{Institute for Cross-Disciplinary Physics and Complex Systems IFISC (UIB-CSIC), E-07122 Palma de Mallorca, Spain}
	
	\author{Rosa L\'opez}
	\affiliation{Institute for Cross-Disciplinary Physics and Complex Systems IFISC (UIB-CSIC), E-07122 Palma de Mallorca, Spain}
	
	\author{Mykhailo Moskalets}
	\affiliation{Institute for Cross-Disciplinary Physics and Complex Systems IFISC (UIB-CSIC), E-07122 Palma de Mallorca, Spain}
	\affiliation{Department of Metal and Semiconductor Physics\char`,~NTU ``Kharkiv Polytechnic Institute'', 61002 Kharkiv, Ukraine}
	
	\author{Sungguen Ryu}
	\email{sungguen@ifisc.uib-csic.es}
	\affiliation{Institute for Cross-Disciplinary Physics and Complex Systems IFISC (UIB-CSIC), E-07122 Palma de Mallorca, Spain}
	
	\date{\today}
	
	\begin{abstract}
		We establish an accessible lower bound on current noise in coherent electron transport, expressed in terms of
		responses to reservoir temperatures and chemical potentials. 
		Derived within the Floquet-Nambu scattering formalism using a Cram\'er--Rao-type inequality, it holds for
		arbitrary linear combinations of charge and heat currents in normal terminals, including those of hybrid superconducting systems, under static or periodic driving.
		The bound is tight in experimentally relevant regimes, notably in quantum point contacts with static or oscillating voltages and normal-superconductor junctions, often providing stronger constraints
		than thermodynamic uncertainty relations.
		It can be evaluated via experimentally accessible response coefficients, without inferring
		entropy production or dynamical activity, and provides useful constraints on the precision of nanoscale thermal and quantum devices.
\end{abstract}

\maketitle

\emph{Introduction.---}
What fundamentally constrains current noise in mesoscopic systems,
where thermal and quantum fluctuations play a significant
role~\cite{blanter2000}?
Can thermal machines deliver power or quantum pumps generate
currents with arbitrarily small relative fluctuations?
At thermal equilibrium, the fluctuation-dissipation theorem relates
fluctuations to linear response~\cite{callen1951,kubo1957}.
Beyond equilibrium, fluctuation bounds constrain current noise
through thermodynamic quantities, dynamical activity, or response
coefficients.

The thermodynamic uncertainty relation (TUR), originally established for stationary classical Markov jump processes \cite{barato2015,gingrich2016}, bounds the signal-to-noise ratio of a current by half the entropy production rate in units of $k_B$. This constraint leads to a trade-off between the efficiency and power precision of steady-state thermal machines \cite{pietzonka2018}.

The applicability of the original TUR beyond classical steady-state dynamics has been extensively investigated. In coherent electronic transport, quantum statistics and energy-selective transmission can lead to violations of the original TUR \cite{Agarwalla2018, timpanaro2025},  
while subsequent works have extended the classical TUR to the quantum regime through modified formulations~\cite{brandner2025,sobrino2026}.
Related studies have explored TURs for heat currents~\cite{Saryal2019}, quantum thermal machines~\cite{rignon-bret2021}, and superconducting hybrid structures \cite{mayo2025,vidal2026}. Periodically driven systems exhibit different precision-dissipation trade-offs, motivating generalized TURs for classical stochastic dynamics and quantum systems \cite{barato2016,koyuk2019,koyuk2020,miller2021,potanina2021}.

Other approaches have also established fundamental limits on current fluctuations. Kinetic uncertainty relations (KURs), originally developed for classical Markov jump processes, constrain current precision through dynamical activity, quantified by the mean rate of stochastic transitions \cite{pietzonka2016,garrahan2017,DiTerlizzi2019KUR}. Unlike entropy production, which characterizes irreversibility, dynamical activity captures the time-symmetric aspect of stochastic dynamics \cite{Maes2020Frenesy}. KURs can provide tighter bounds than TURs far from equilibrium. Quantum coherence can invalidate classical KURs, motivating generalized kinetic bounds \cite{prech2025,blasi2026,palmqvist2026}. Further uncertainty relations have been derived from fluctuation theorems \cite{hasegawa2019FT,timpanaro2019,potts2019,tesser2026}, or by combining thermodynamic and kinetic constraints \cite{vo2022}.

More directly, information-theoretic inequalities involving Fisher
information and dynamical responses establish general links between
fluctuations and response~\cite{hasegawa2019,dechant2020}.
Response TURs and KURs constrain the precision of current responses
through entropy production or dynamical activity in Markov 	processes~\cite{ptaszynski2024,aslyamov2025}. In coherent electronic transport, out-of-equilibrium
fluctuation-dissipation bounds constrain charge current noise in terms of
average currents and nonequilibrium conditions~\cite{tesser2024FDB}, with extensions to periodically
driven conductors~\cite{tesser2026DrivenFDB}. 

Despite these advances, existing bounds often rely on specific dynamical assumptions, are formulated for particular types of current, or involve entropy production, dynamical activity, or information-theoretic quantities~\cite{funo2015}, which may not be directly accessible from standard transport measurements.
This motivates a search for broadly applicable bounds on current noise expressed directly in terms of experimentally accessible response coefficients.

In this work, we establish a thermo-electric response bound, valid for quantum coherent conducting systems, that constrains current noise in terms of differential response coefficients with respect to reservoir chemical potentials and temperatures.
For charge currents the bound reads 
\begin{equation}
	S \ge \sum_N k_B T_N \frac{h}{e^2} 
	\left( G_N^2 + \frac{L_N^2}{ \mathcal{L}_0 } \right) ,
	\label{eq1}
\end{equation}
where $S$
denotes the zero-frequency 
charge current noise. 
Here, $G_N $ and $L_N $ are, respectively, the differential electrical and thermoelectric conductances with $N$ labeling the perturbed normal terminal, and $\mathcal{L}_0=\pi^2k_B^2/(3e^2)$ is the Sommerfeld value of the Lorenz number. See Eq.~\eqref{eq:optimized_bound} for linear combinations of charge and heat currents.
We derive this bound within the
Floquet-Nambu scattering formalism~\cite{moskalets2002,burset2025}
using a Cram\'er-Rao-type  inequality~\cite{rao1945}.
The bound holds for 
arbitrary configurations of coherent conductors, including hybrid superconducting structures and under arbitrary periodic driving.

\emph{Setup.---} 
We consider transport through a periodically driven quantum conductor with driving angular frequency \(\Omega\).
We use a quasienergy
\(\varepsilon\in[0,\hbar\Omega)\) and define
\(E_n=\varepsilon+n\hbar\Omega .\)
A propagating Floquet-Nambu state is labeled by
\( a=(K,\alpha,n), \)
where \(K\) labels the terminals,
\(\alpha=\pm1\) the Nambu sectors, and \(n\)  the Floquet sidebands.
We assume a single transverse mode per state, but a generalization is straightforward.
We take \(\alpha=-1\) for electron-like states and \(\alpha=+1\) for hole-like
states, consistently with the charge weight used below.
At fixed quasienergy, incoming and outgoing operators are related
by a unitary scattering matrix in the propagating extended space,
\(b_a(\varepsilon)=\sum_b \mathcal{S}_{ab}(\varepsilon)a_b(\varepsilon).\)
All superconducting phases, gaps, and time-dependent parameters are encoded in
\(\mathcal{S}\).
We introduce diagonal matrices
\(\hat f_{ab}(\varepsilon)=\delta_{ab}f_a(\varepsilon) \),
where for a normal terminal
\( f_a(\varepsilon) = f_K^\alpha(E_n)
=
[
\exp\{
(E_n+\alpha \mu_K)/(k_B T_K)\}+1
]^{-1}
\)
is the Fermi occupation,
\(\delta_{ab}\) denotes the equality of all
components of the composite label, and
\(T_K\) is the temperature of the terminal $K$.

\emph{Currents and noise.---}
Let \(X\) denote the transported quantity, namely charge or heat. 
The corresponding factors for the current carried by a propagating quasiparticle
of type \(\alpha\) in a normal terminal \(N\) are
\( x_N^{C,\alpha}=e\alpha \)
and
\( x_N^{Q,\alpha} (E)=E+\alpha\mu_N \).
We introduce a diagonal matrix
\(\left(\mathcal{X}_N^X\right)_{ab}
=
\delta_{ab}\delta_{KN}\,
x_N^{X,\alpha}(E_n) \).
The time-averaged current flowing out of terminal \(N\) is
\begin{equation}
	J_N^X
	=
	\frac{1}{2h}
	\int_0^{\hbar\Omega}d\varepsilon\,
	\operatorname{Tr}\left[
	\hat f\,\mathcal A_N^X
	\right],
	\label{eq:J_def_draft}
\end{equation}
where $\mathcal A_N^X$ is the transport matrix
\( \mathcal A_N^X=\mathcal X_N^X-\mathcal S^\dagger \mathcal X_N^X \mathcal S \).
The factor \(1/2\) compensates for Nambu double counting, following the
charge-current normalization of Ref.~\cite{burset2025}.
The trace is over the extended channel-Nambu-Floquet space at fixed
\(\varepsilon\).
In the absence of periodic driving, the integral from $0$ to $\hbar\Omega$ with respect to the quasienergy and the trace over the Floquet sidebands combine to the energy integral from $-\infty$ to $\infty$.

For arbitrary matrices \(B\) and \(C\) in the same space, define an inner product
\(
\langle B,C\rangle_f
=
\int_0^{\hbar\Omega}d\varepsilon
\sum_{a,b}
f_a(1-f_b)B_{ab}C_{ab}^* /(2h).
\)
Because the weights \(f_a(1-f_b)\) are non-negative, this product is positive
semidefinite and obeys the Cauchy-Schwarz inequality, 
\( |\braket{B,C}_f|^2 \le \braket{B,B}_f \braket{C,C}_f\),
see Supplemental Material (SM) for details.
The zero-frequency correlations can then be
written as
\(S_{KL}^{XY}
=
\langle
\mathcal A_K^X,\mathcal A_L^Y
\rangle_f \).
Thus, the autocorrelation of any real linear combination of currents
\(J=\sum_{N,X}w_N^XJ_N^X \)
is a squared norm,
\begin{equation}
	S=\langle\mathcal A,\mathcal A\rangle_f ,
	\label{eq:SJ_def_draft}
\end{equation}
where
\(\mathcal A=\sum_{N,X}w_N^X\mathcal A_N^X \).
$S$ is non-negative as it is a squared norm.

\emph{Response coefficients.---}
Our approach is inspired by the TUR for periodically driven Markov systems~\cite{koyuk2019}, whose derivation uses an auxiliary perturbation that modifies the asymmetry between forward and backward transition rates. 
In coherent scattering, varying reservoir temperatures and chemical potentials provides a simple and natural way to modify the directional bias of transport processes. We describe such variations in the normal terminals by the directional derivative
\(\mathcal D=\sum_N(\Theta_N\partial_{T_N}+\Lambda_N\partial_{\mu_N})\),
where \(\Theta_N\) and \(\Lambda_N\) are arbitrary real coefficients specifying a direction in the space of reservoir parameters. The response $\mathcal{D} J$ has two contributions:
$ 	\int_0^{\hbar\Omega} \! d\varepsilon
\operatorname{Tr} [ (\mathcal{D} \hat f) \mathcal A]/(2h) $
and
$ 	\int_0^{\hbar\Omega} \! d\varepsilon
\operatorname{Tr} [\hat f\,(\mathcal{D} \mathcal A)] /(2h) 	$.
For noninteracting carriers, where the scattering matrix is held fixed, 
the second contribution then becomes
$ \sum_N \Lambda_N w^Q_N J^C_N/e $, since only the heat factor depends explicitly on the reservoir parameters, through $\partial_{\mu_N}x_N^{Q,\alpha}=\alpha=x_N^{C,\alpha}/e$.
Applying the properties of the Fermi functions to the first contribution (see SM), the response becomes
\begin{equation}
	\mathcal D J
	=
	\langle \mathcal A, \mathcal{H}\rangle_f
	+\sum_N \Lambda_N w^Q_N J^C_N/e.
	\label{eq:D_inner}
\end{equation}
We have introduced a diagonal Fermi-response matrix which encodes the response of the Fermi occupations, with diagonal entries
\( \mathcal{H}_{aa}
\equiv (\mathcal{D}f_a)/[ f_a(1-f_a)]=
[
\Theta_K
(E_n+\alpha\mu_K)/(k_B T_K^2)
- \alpha\Lambda_K/(k_B T_K)
]
\) 
for normal terminals $K$ and $\mathcal{H}_{aa}=0$ otherwise.

\emph{Bound.---}
Applying the Cauchy-Schwarz inequality to the inner product in Eq.~\eqref{eq:D_inner} gives
\begin{equation}
	\Big(\mathcal D J
	-\sum_N \Lambda_N w^Q_N J^C_N/e
	\Big)^2
	\le
	S\, \langle \mathcal{H},\mathcal{H}\rangle_f .
	\label{eq:directional_bound}
\end{equation}
This equation has the fluctuation-response structure of the Cram\'er-Rao inequality~\cite{rao1945}. 
The left-hand side is the squared directional response, with the contribution from the explicit chemical-potential dependence of the heat-current factors \(x_N^{Q,\alpha}(E)=E+\alpha\mu_N \) subtracted. The factor $\braket{\mathcal{H}, \mathcal{H}}$ is the directional Fisher information rate about reservoir parameters encoded in the fermionic occupations (see SM).
Using the Fermi identities, the rate reduces to
\begin{equation}
	\langle \mathcal{H},\mathcal{H}\rangle_f
	=
	\sum_{N}
	\frac{1}{h k_B T_N}
	\left(
	\ell_0 k_B^2 \Theta_N^2+\Lambda_N^2
	\right).
	\label{eq:norm-H}
\end{equation}
$\ell_0=\pi^2/3$ is the Sommerfeld value of the dimensionless Lorenz number. 
Eqs.~\eqref{eq:directional_bound} and \eqref{eq:norm-H} impose a lower bound for the noise $S$.
Since the bound holds for every auxiliary
direction \(\{\Theta_K,\Lambda_K\} \), we choose the one that maximizes it.
The optimization is equivalent to that in the multiparameter Cram\'er-Rao bound, and in our case dictates the optimal 
$\mathcal{H}$ to be most aligned with $\mathcal{A}$. 
This leads to the optimal direction, $\Theta_N = T_N \partial_{T_N} J/\ell_0 $ and  
$\Lambda_N = k_B^2 T_N (\partial_{\mu_N} J - w_N^Q J_N^C/e) $
up to an overall constant, (see SM) and our main result,
\begin{equation}
	S
	\ge
	\sum_{N}
	h k_B T_N
	\left[
	\frac{1}{\ell_0 k_B^2}
	\Big( \frac{\partial J}{\partial T_N} \Big)^2
	+ \Big(\frac{\partial J}{\partial \mu_N}-\frac{w^Q_N}{e} J^C_N \Big)^2
	\right] .
	\label{eq:optimized_bound}
\end{equation}
In the case of charge current, it reduces to Eq.~\eqref{eq1}.
Retaining any subset of the nonnegative terms on the right-hand side also yields a valid noise bound, which may be useful when only some response coefficients are experimentally accessible.
We refer to Eq.~\eqref{eq:optimized_bound} as a thermo-electric response bound (TRB). 

The TRB is saturated when $\mathcal{A} $ belongs to the span of all Fermi-response matrices $\mathcal{H} $, in the inner-product space defined by $\braket{\cdot,\cdot}_f$. 
As these are diagonal, the TRB cannot capture the off-diagonal contribution of the noise, $S_{\text{off}}=\int_0^{\hbar \Omega} d\varepsilon \sum_{a\neq b} f_a(\varepsilon) (1-f_b(\varepsilon)) |\mathcal{A}_{ab}|^2/(2h) $, associated with partitioning by the scatterer. 
When the normal-channel diagonal contribution dominates the noise, the TRB is tight if the diagonal entries of transport matrix are approximately affine in energy, so that they are well described by Fermi-response matrices, see below for examples. 
Note that the TRB does not guarantee the saturation for the equilibrium noise, unlike the classical and quantum TURs, because the off-diagonal contribution does not vanish generally.

\emph{Perfect-Andreev saturation.---}
Consider a
normal terminal with a single transverse mode, coupled to a grounded
superconductor. The bound is saturated for perfect Andreev reflection throughout the transport window. A transparent normal-superconductor (NS) junction approaches this limit for \(k_BT_N,|\mu_N|\ll\Delta\ll\epsilon_F\) where $\epsilon_F$ is the Fermi energy.
For the charge current out of the normal terminal,
\(\mathcal X_N^C=e\,\tau_z\)
in the Nambu basis
\(\tau_z=\operatorname{diag}(-1,+1)\).
Thus,
\( \mathcal S^\dagger \mathcal X_N^C \mathcal S
=
-e \tau_z  \)
and
\(\mathcal A
=
2e\,\tau_z  \).
When perturbing only the chemical potential of the normal lead,
\(\mathcal{H} =
-\tau_z/(k_B T) \)
which is proportional to $\mathcal{A}$.
Hence, the Cauchy-Schwarz inequality is saturated.
Explicitly,
\(J=-2e\mu_N/{h} \) and
\(S=4e^2 k_B T/{h} \).
Hence \( \partial J/\partial T_N =0\), \(\partial J/\partial\mu_N =-2e/h\), and Eq.~\eqref{eq:optimized_bound} is saturated. 
The factor \(2e\) in $J$ is the transferred Cooper-pair charge.

\begin{figure}[t]
	\centering
	\includegraphics[width=1\linewidth]{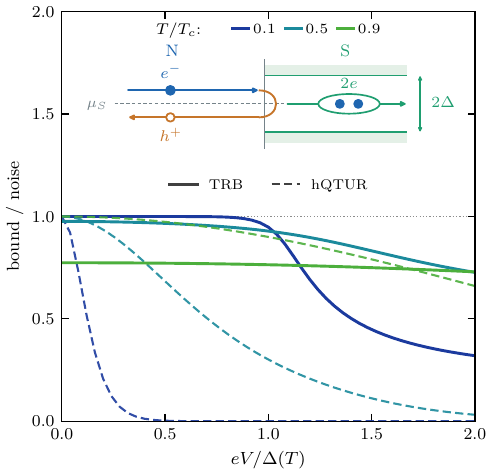}
	\caption{TRB in a finite-gap NS junction. 
		The normal terminal is electrically biased by \(V\), while the superconducting terminal, with critical temperature $T_c$ and gap $\Delta(T)$, is grounded. Both terminals are at temperature \(T\). 
		Solid curves show the tightness of the TRB for the charge current, namely the ratio between the two sides of Eq.~\eqref{eq:optimized_bound}.
		The TRB is tight when the transport is dominated by Andreev reflection, namely at low temperatures and subgap biases. 
		Dashed curves show the tightness of the hybrid quantum TUR. 
	}
	\label{fig:NS}
\end{figure}

\emph{Finite-gap NS.---}
We now test the TRB in a finite-gap normal-superconducting
junction. The setup (see the inset of Fig.~\ref{fig:NS}) consists of one biased normal lead connected to a grounded
superconducting terminal. 
Both terminals are at temperature $T_S=T_N=T$.
The scattering matrix is computed from the
Bogoliubov-de Gennes matching problem at the NS interface beyond the Andreev
approximation~\cite{Blasi2023Hybrid}.  The full energy dependence of the electron, hole, and
superconducting quasiparticle wave vectors is retained, rather than replacing
them by their Fermi-surface values. 
The explicit finite-gap scattering matrix
and amplitudes used in the calculation are summarized in SM.
The superconducting gap entering these
amplitudes is temperature dependent and follows the interpolation
\(\Delta(T)=\Delta(0)\tanh[1.74\sqrt{T_c/T-1}]\), with \(\Delta(T)=0\) for
\(T\ge T_c\). This makes the test sensitive both to electron-hole conversion and
to the explicit temperature dependence of the hybrid scattering problem. 
When evaluating \(\partial_{T_N}J\) in the TRB, we hold \(T_S\), and hence the superconducting gap, fixed.
Here we focus on the charge current flowing out of the normal terminal,
denoted by \(J\equiv J_N^C\), and on its zero-frequency charge noise
\(S\equiv S_{NN}^{CC}\).

Figure~\ref{fig:NS} shows the TRB as a function of the reduced bias \(eV/\Delta(T)\).
Below the gap edge, especially at low temperature, transport is dominated by Andreev
reflection and the TRB is tight.
Once \(eV\) exceeds the gap $\Delta(T)$, quasiparticle channels in the superconducting terminal become available.
Partitioning between Andreev reflection and quasiparticle transmission contributes to the off-diagonal noise $S_\textrm{off}$, reducing the tightness of the TRB.
As the temperature approaches $T_c$, the thermally excited quasiparticles in the superconductor also produce a diagonal noise contribution that is not captured by the TRB, making the bound loose. 
We compare our bound with hybrid quantum TUR for charge currents~\cite{mayo2025,vidal2026} which reads
\begin{equation}
	S \ge 2eJ \operatorname{csch} \left( \frac{ \sigma e}{k_B J}\right) ,
	\label{eq:hqTUR}
\end{equation}
where $\sigma= -(J_N^Q+J_S^Q)/T$ is the entropy production rate.
The comparison (solid vs dashed lines) shows that the TRB provides tighter bounds when the temperature is much smaller than the critical temperature or when the electrical bias is much larger than the gap.

\begin{figure}[t]
	\centering
	\includegraphics[width=1\linewidth]{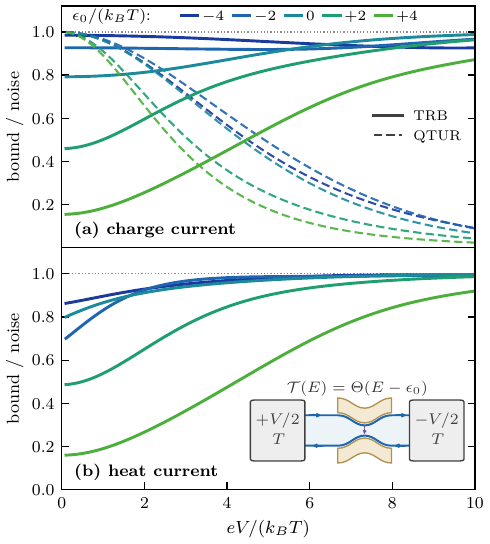}
	\caption{TRB for quantum point contacts. Solid lines show the tightness of the TRB for (a) the charge current (blue to green as the barrier height $\epsilon_0$ increases), and (b) the heat current, as a function of the electrical bias $V$. 
		The TRB is fairly tight unless the barrier lies well above both chemical potentials. 
		Dashed lines in (a) show the tightness of the quantum TUR. Parameters: $T_L=T_R=T$,  step transmission, and $\epsilon_0$ is measured with respect to the average chemical potential. 
	}
	\label{fig:step}
\end{figure}

\emph{Quantum point contacts.---}
When the terminals are normal, we find that our bound is often tight for transport through quantum point contacts.
Figure~\ref{fig:step} shows the result for the charge and heat currents out of the right terminal (see the inset for the setup).
The reservoirs are isothermal, namely $T_L=T_R=T$, and electrically biased as $\mu_L = \bar{\mu} + eV/2$ and $\mu_R=\bar{\mu}- eV/2$.
The transmission is a step function,
namely $\mathcal{T}(E) = 1 $ for $E> \epsilon_0$ and 0 otherwise.
A single transverse mode is assumed.
The TRB (solid lines) is fairly tight unless the barrier lies well above both chemical potentials, namely unless $\epsilon_0 -\max(\mu_L, \mu_R)\gg  k_B T$.
We compare our bound with the quantum-TUR bound for charge current~\cite{brandner2025}, which reads as
$  S \ge eJ \operatorname{csch}[  \sigma e/(2k_B J)]  $, which follows from Eq.~\eqref{eq:hqTUR} by replacing $e$ with $e/2$ and using $\sigma=- (J_L^Q+J_R^Q)/T $.
For the parameters shown, the TRB is tighter at sufficiently large bias, while the QTUR is tighter near equilibrium.

For the step transmission, scattering is deterministic at
every energy, so $S_{\mathrm{off}}=0$. Tightness is therefore
determined entirely by how well the diagonal transport matrix
is represented by the Fermi-response matrices.
When the barrier lies well below both chemical potentials, 
the diagonal entries are approximately constant (for charge current) or linear (for heat current) 
over both thermal windows, making the TRB tight. 
When the barrier lies well above both chemical potentials,
the nonzero entries are confined to the thermal tails.
This profile is poorly represented by the response
matrices, making the TRB loose.
In the intermediate regimes, 
the diagonal transport matrix can retain a substantial projection on to the span of the response matrices.  
For example, at $\epsilon_0=\mu_L=\mu_R$ and for charge current, 
\(
\mathcal A=e\,\Theta(x)\operatorname{diag}(1,-1)
\)
in the reduced lead basis $(L,R)$
where $x=(E-\mu)/(k_BT)$.
The optimal response matrix is
\(
\mathcal H_{\mathrm{opt}}\propto
[1/2+(\ln2/\ell_0)x]\operatorname{diag}(1,-1).
\)
Their squared normalized overlap is approximately $80\%$, explaining
the fairly tight TRB despite the sharp transmission threshold.
At large positive bias, with
\(\mu_L-\epsilon_0,\epsilon_0-\mu_R\gg k_BT\),
the charge-current transport matrix satisfies
\(\mathcal A\simeq e\,\operatorname{diag}(1,0)\)
in the inner product space.
Varying only \(\mu_L\) gives
\(\mathcal H\propto\operatorname{diag}(1,0)\),
so the TRB approaches saturation.
The same tightness is obtained for the heat current.
See SM for other transmission profiles.

\begin{figure}[t]
	\centering
	\includegraphics[width=1\linewidth]{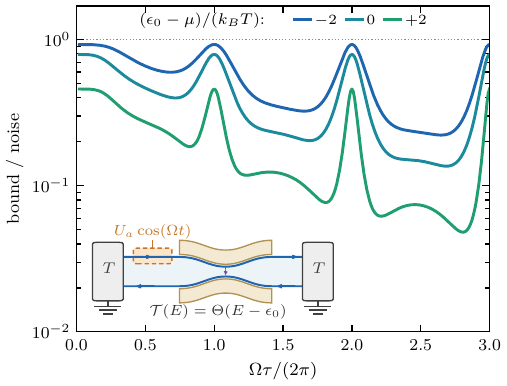}
	\caption{TRB for charge current pumped through a quantum point contact by AC voltage. 
		The TRB is fairly tight in the adiabatic regime unless
		$\epsilon_0-\mu\gg k_BT$.
		In the nonadiabatic regime, photoassisted partitioning reduces the tightness of the TRB, except at integer values of $\Omega\tau/(2\pi)$, where the AC-induced phase vanishes. 
		Parameters: $U_\text{a}= \hbar\Omega$,  $\mu_L=\mu_R=\mu$, and $T_L=T_R=T=\hbar/(k_B \tau) $ where $\tau$ is travel time in the AC region.
	}
	\label{fig:driven_step}
\end{figure}

{\it Periodically driven systems.---}
We consider the same quantum point contact, with AC potential $U(t) = U_\text{a} \cos \Omega t$ applied to a finite region of the left lead with travel time $\tau$ (see the inset of Fig.~\ref{fig:driven_step}).
This modulation generates a charge current through photoassisted tunneling~\cite{platero2004, dashti2019,ryu2021}.
Figure~\ref{fig:driven_step} 
shows the TRB for the pumping current. 
The TRB is fairly tight in the adiabatic regime, $\Omega \tau \ll 2\pi $, unless the barrier lies well above the chemical potential, as in the static case. 
As the frequency increases, photon absorption and emission in the AC region can induce partitioning between transmission and reflection, contributing to the off-diagonal noise $S_\textrm{off}$.
Consequently, the tightness of TRB is reduced, as reflected in the overall decreasing trend of the curves. At integer values of $\Omega \tau/(2\pi)$, the phase accumulation by the AC voltage during the travel time vanishes. The transport then reduces to the static case: the mean pumped current vanishes, and the bound recovers its static tightness. 
Away from these points, the TRB remains informative for nonzero pumped currents in the nonadiabatic regime, providing a lower bound within roughly one order of magnitude of the actual noise for the parameters shown. 


The TRB can serve as a useful benchmark for pumping precision without direct noise measurements.
Characterizing strongly suppressed pumping fluctuations through noise measurements~\cite{robinson2005} or single-electron counting~\cite{fricke2014} can be experimentally demanding: resolving rare transfer errors requires extensive sampling~\cite{giblin2016}, while weak intrinsic current noise must be distinguished from the measurement background~\cite{maire2008}.
For quantum pumps used in current metrology~\cite{pekola2013}, the TRB identifies stability against reservoir perturbations as a cost of precision. Achieving low noise at a prescribed mean current, typically corresponding to one electron transferred per driving period, requires the operating point to lie on a locally stable plateau~\cite{maire2008} of the mean current with respect to reservoir temperatures and chemical potentials. This requirement can help rule out operating points that cannot meet a prescribed precision target and guide the choice of pumping parameters.

\emph{Discussion.---}
The TRB derived here constrains current noise through experimentally accessible reservoir responses and applies to arbitrary linear combinations of currents measured in normal terminals under arbitrary periodic driving. 
Calculations demonstrate that our bound is fairly tight at low temperatures when the scattering is deterministic and energy-selective, unless all Fermi edges lie well outside the transmitting energy windows. The experimental accessibility and tightness of the TRB support its application to nanoscale thermoelectrics~\cite{whitney2014,benenti2017, whitney2018, chrirou2026, Ozaeta2014ThermoNIS, giazotto2015, Kolenda2016SeebeckFS, Arrachea2025ThermoNS,Dutta2025ThermalSC,Blasi2021Engines} and quantum pumps~\cite{kaestner2015,ryu2016,yamahata2019}.

For charge currents, the inequalities \((\partial_{\mu_N}J)^2/S\le1/(hk_BT_N)\) and \((\partial_{T_N}J)^2/S\le \ell_0k_B/(hT_N)\) share the fluctuation–response structure of response TURs and KURs~\cite{ptaszynski2024,aslyamov2025}. The cited bounds for stationary Markov processes involve transition-rate perturbations and entropy production or dynamical activity, while the quantum susceptibility-KUR involves responses to the system–reservoir coupling and a partial dynamical activity~\cite{palmqvist2026}. 
In contrast, in the present bound, the coefficients are fixed by reservoir Fermi statistics, without requiring either dynamical cost.

Above we have assumed noninteracting carriers.
For weakly interacting carriers, variations in reservoir voltages or temperatures can induce charge redistribution that modifies the scattering matrix through screening~\cite{christen1996,sanchez2013a}.
Strong capacitive coupling to a gate held at fixed voltage can suppress changes in the scatterer potential during reservoir perturbations. If the scattering matrix stays effectively fixed, the measured responses approximate those in Eq.~\eqref{eq:D_inner}.

\emph{Acknowledgment.---}
The authors acknowledge support by  the Spanish State Research Agency (MCIN/AEI/10.13039/501100011033),
FEDER (UE) under Grant No. PID2023-151975NB-I00.
M. M. acknowledges the support from CSIC/IUCRAN2022 under Grant No. UCRAN20029.

\bibliography{reference}

\clearpage

\setcounter{equation}{0}
\renewcommand{\theequation}{S\arabic{equation}}
\setcounter{figure}{0}
\renewcommand{\thefigure}{S\arabic{figure}}
\renewcommand{\theHequation}{supp.\arabic{equation}}
\renewcommand{\theHfigure}{supp.\arabic{figure}}

\onecolumngrid
\begin{center}
	{\bfseries\large Supplemental Material:\\ Thermo-electric Response Bound on Fluctuations in Coherent Transport}
\end{center}

\twocolumngrid

\section{Weighted Scalar Product}
\label{app:inner_product}

The product $\braket{B,C}_f$ introduced in the main text is the natural quadratic
form associated with fermionic zero-frequency noise. In compact
matrix notation it reads
\begin{equation}
	\langle B,C\rangle_f
	=
	\frac{1}{2h}
	\int_0^{\hbar\Omega}d\varepsilon\,
	\operatorname{Tr}\left[
	\hat f B(1-\hat f)C^\dagger
	\right].
	\label{eq:inner_trace_appendix}
\end{equation}
Expanding the trace gives
\[
\operatorname{Tr}\left[
\hat f B(1-\hat f)C^\dagger
\right]
=
\sum_{a,b}f_a(1-f_b)B_{ab}C_{ab}^*,
\]
which is the definition given in the main text. The factor \(f_a(1-f_b)\)
has the standard fermionic interpretation: the incoming state \(a\) must be
occupied while the incoming state \(b\) must be available. Since
\(0\le f_a\le1\),
\begin{equation}
	\langle B,B\rangle_f
	=
	\frac{1}{2h}
	\int_0^{\hbar\Omega}d\varepsilon
	\sum_{a,b}
	f_a(1-f_b)|B_{ab}|^2
	\ge0 .
	\label{eq:inner_positive_appendix}
\end{equation}
Equivalently, with
\(\widetilde B=\hat f^{1/2}B(1-\hat f)^{1/2}\),
\[
\langle B,B\rangle_f
=
\frac{1}{2h}
\int_0^{\hbar\Omega}d\varepsilon\,
\operatorname{Tr}\left[\widetilde B\widetilde B^\dagger\right].
\]
Therefore the Cauchy-Schwarz inequality applies. Substituting the transport
matrices \(\mathcal A_K^X\) gives the zero-frequency correlations \(S_{KL}^{XY}\) of the main text; for a real linear combination of currents this is
the autocorrelation of the measured signal and hence a genuine norm.

\section{Fisher information in Fermi-Dirac occupations}

Consider a fermionic reservoir with temperature $T$ and chemical potential $\mu$. 
It is described by a set of independent Bernoulli variables $\{n(E)\}$,  where $n(E)$ describes the occupation of a state at energy $E$ with success probability  $f(E) = 1/[1+\exp( (E-\mu)/(k_BT))]$.  
Using that the Fisher information about success probability $p$ of a Bernoulli variable is $p^{-1} +(1-p)^{-1}$,
we obtain the Fisher information rate matrix about the reservoir parameters contained in the occupations,
\begin{align}
	I &= \text{diag} (I_T,  I_\mu )  , \\
	I_{T}
	&= \int_{-\infty}^\infty \frac{dE}{h} \left(\frac{E-\mu}{k_B T^2}\right)^2 f(E) \big(1-f(E) \big) ,\\
	I_\mu  &= \int_{-\infty}^\infty \frac{dE}{h}  \frac{1}{(k_B T)^2} f(E) \big(1-f(E) \big) .
\end{align}
Here, \(dE/h\) counts incoming states per unit time in the standard wave-packet description; accordingly, \(I\) is a Fisher information rate matrix.
The off-diagonal element,  $\braket{ (\partial_T \ln P)(\partial_\mu \ln P ) } $, where $P$ denotes the joint probability distribution of the Bernoulli variables and $\braket{\cdots}$ denotes the corresponding mean value, vanishes due to the energy integral of an odd function.

Now consider all the normal terminals. They are described by a set of independent Bernoulli variables $\{n_N(E)\}$.
The Fisher information rate matrix becomes 
\begin{align}
	\mathcal{I} 
	&= \mathcal{I}_T \oplus \mathcal{I}_\mu ,\\
	\mathcal{I}_{T}
	&= \bigoplus_N I_{T_N}  ,\\
	\mathcal{I}_{\mu}
	&= \bigoplus_N I_{\mu_N} .
\end{align}

With these definitions, the norm $\braket{\mathcal{H}, \mathcal{H}}_f$ in the main text, see Eqs.~\eqref{eq:directional_bound} and \eqref{eq:norm-H}, can be interpreted as directional Fisher information rate
\begin{equation}
	\braket{\mathcal{H}, \mathcal{H}}_f
	= \sum_N \left[
	\Theta_N^2 (\mathcal{I}_T )_{NN}
	+\Lambda_N^2 (\mathcal{I}_\mu )_{NN} \right] .
\end{equation}

Evaluating the energy integrals gives
\begin{align}
	\mathcal{I}_T
	&= \bigoplus_K \frac{\pi^2 k_B}{3h  T_K }   ,  \\     \mathcal{I}_\mu
	&= \bigoplus_K \frac{1}{ hk_B T_K }    . 
\end{align}

\section{Fermi Integrals}
\label{app:fermi_integrals}

The evaluation of Eq.~\eqref{eq:norm-H} uses only the Nambu pair of
Fermi functions. Let
\[
g(x)=\frac{e^x}{(e^x+1)^2}
=\frac{1}{4\cosh^2(x/2)} .
\]
Changing variables to \(x=(E+\alpha\mu_K)/(k_B T_K) \), the two Nambu branches yield identical
full-axis integrals, whose sum is compensated by the factor \(1/2\).
Since \(g(x)\) is even,
\begin{align*}
	\int_{-\infty}^{\infty}dx\,g(x)&=1,\\
	\int_{-\infty}^{\infty}dx\,xg(x)&=0,\\
	\int_{-\infty}^{\infty}dx\,x^2g(x)&=\frac{\pi^2}{3}.
\end{align*}
It follows that
\begin{equation}
	\frac12\int_{-\infty}^{\infty} dE
	\sum_{\alpha=\pm1}
	f_K^\alpha(1-f_K^\alpha)
	=
	k_B T_K,
	\label{eq:I0_appendix}
\end{equation}
\begin{equation}
	\frac12\int_{-\infty}^{\infty} dE
	\sum_{\alpha=\pm1}
	\alpha(E+\alpha\mu_K)f_K^\alpha(1-f_K^\alpha)
	=0,
	\label{eq:I1_appendix}
\end{equation}
and
\begin{equation}
	\frac12\int_{-\infty}^{\infty} dE
	\sum_{\alpha=\pm1}
	(E+\alpha\mu_K)^2f_K^\alpha(1-f_K^\alpha)
	=
	\frac{\pi^2}{3} k_B^3T_K^3 .
	\label{eq:I2_appendix}
\end{equation}

\section{Optimization over the Direction}
\label{app:direction_optimization}

Write the left-hand side of Eq.~\eqref{eq:directional_bound} as
\(\big[\sum_N(\Theta_Na_N+\Lambda_Nb_N)\big]^2\), with
\begin{equation}
	a_N=\frac{\partial J}{\partial T_N},
	\qquad
	b_N=\frac{\partial J}{\partial\mu_N}-\frac{w_N^Q}{e}J_N^C .
	\label{eq:ab_appendix}
\end{equation}
The optimization is a diagonal Rayleigh quotient. Introduce the rescaled variables 
\begin{equation}
	u_{T_N}
	=
	\sqrt{\frac{\ell_0 k_B}{h T_N}}\,\Theta_N,
	\qquad
	u_{\mu_N}
	=
	\sqrt{\frac{1}{h k_B T_N}}\,\Lambda_N ,
	\label{eq:u_rescaling_appendix}
\end{equation}
and
\begin{equation}
	v_{T_N}
	=
	\sqrt{\frac{h T_N}{\ell_0 k_B}}\,a_N,
	\qquad
	v_{\mu_N}
	=
	\sqrt{h k_B T_N}\,b_N .
	\label{eq:v_rescaling_appendix}
\end{equation}
Let \(\boldsymbol u\) and \(\boldsymbol v\) be the Euclidean vectors with
components \(u_{T_N},u_{\mu_N}\) and \(v_{T_N},v_{\mu_N}\). By
Eq.~\eqref{eq:norm-H}, \(\langle H,H\rangle_f=\boldsymbol u\cdot\boldsymbol u\), and the
left-hand side of Eq.~\eqref{eq:directional_bound} is
\((\boldsymbol u\cdot\boldsymbol v)^2\). Thus, for any nonzero direction,
Cauchy-Schwarz gives
\begin{equation}
	\frac{(\boldsymbol u\cdot\boldsymbol v)^2}
	{\boldsymbol u\cdot\boldsymbol u}
	\le
	\boldsymbol v\cdot\boldsymbol v
	=
	\sum_{N}
	h k_B T_N
	\left[
	\frac{1}{\ell_0 k_B^2}a_N^2+b_N^2
	\right].
	\label{eq:direction_optimization_appendix}
\end{equation}
The equality condition is \(\boldsymbol u\parallel\boldsymbol v\). Since
Eq.~\eqref{eq:directional_bound} holds for arbitrary
\(\Theta_N,\Lambda_N\), this parallel direction can be chosen, and
Eq.~\eqref{eq:direction_optimization_appendix} gives the optimized bound in
Eq.~\eqref{eq:optimized_bound}. The maximizing direction is defined up to
an overall scale and can be written as
\begin{equation}
	\Theta_N
	\propto
	\frac{h T_N}{\ell_0 k_B}a_N,
	\qquad
	\Lambda_N
	\propto
	h k_B T_N\,b_N.
	\label{eq:optimal_direction_appendix}
\end{equation}
If all response coefficients \(a_N,b_N\) vanish, the quotient is zero for every direction and
the optimized bound is correspondingly trivial.

\section{Finite-Gap NS Scattering Matrix}
\label{app:ns_scattering}

This appendix gives the finite-gap normal-superconducting scattering model used
for Fig.~\ref{fig:NS}. We consider a single-channel two-terminal NS
junction, with a normal reservoir coupled to a grounded superconducting
reservoir. In the Nambu basis
\((e,h,\tilde e,\tilde h)\), where \(e,h\) denote electron and hole states in
the normal lead and \(\tilde e,\tilde h\) denote electron-like and hole-like
Bogoliubov quasiparticles in the superconducting lead, the scattering matrix is
\begin{equation}
	S_{NS}=
	\begin{pmatrix}
		r_{ee} & r_{eh} & t_{e\tilde e} & t_{e\tilde h}\\
		r_{he} & r_{hh} & t_{h\tilde e} & t_{h\tilde h}\\
		t_{\tilde e e} & t_{\tilde e h} & r_{\tilde e\tilde e} & r_{\tilde e\tilde h}\\
		t_{\tilde h e} & t_{\tilde h h} & r_{\tilde h\tilde e} & r_{\tilde h\tilde h}
	\end{pmatrix}.
	\label{eq:S_NS_appendix}
\end{equation}
The diagonal elements describe normal reflection and quasiparticle reflection,
while the off-diagonal describe Andreev conversion and quasiparticle
transmission between the normal and superconducting sides.
The superconducting
order parameter is finite and temperature dependent. In the numerics we use
\begin{equation}
	\Delta(T)
	=
	\Delta(0)
	\tanh\left[
	1.74\sqrt{\frac{T_c}{T}-1}
	\right],
	\qquad T<T_c,
	\label{eq:Delta_T_appendix}
\end{equation}
and set \(\Delta(T)=0\) for \(T\ge T_c\). 

The interface may include a dimensionless barrier parameter \(Z\). The numerical
code allows finite \(Z\), with \(Z=0\) corresponding to a transparent interface.
The explicit amplitudes below are the transparent-interface coefficients
(\(Z=0\)) obtained by matching the Bogoliubov-de Gennes wavefunctions without
making the Andreev approximation. Thus the electron and hole wave vectors are
not replaced by \(k_F\), and the superconducting quasiparticle wave vectors keep
their full energy dependence.

\subsection{Scattering Coefficients Beyond the Andreev Approximation}
\label{app:ns_coefficients}

Let \(\epsilon\) be the positive excitation energy, \(\epsilon_F\) the Fermi
energy, \(k_F=\sqrt{2m\epsilon_F}/\hbar\), \(x\) the interface coordinate used in
the matching, and \(\phi\) the superconducting phase. The normal electron and
hole wave vectors are
\begin{equation}
	k_e=k_F\sqrt{1+\frac{\epsilon}{\epsilon_F}},
	\qquad
	k_h=k_F\sqrt{1-\frac{\epsilon}{\epsilon_F}},
	\label{eq:ke_kh_appendix}
\end{equation}
while the superconducting electron-like and hole-like quasiparticle wave
vectors are
\begin{equation}
	q_{\tilde e}
	=
	k_F\sqrt{1+\sqrt{\frac{\epsilon^2-\Delta^2}{\epsilon_F^2}}},
	\qquad
	q_{\tilde h}
	=
	k_F\sqrt{1-\sqrt{\frac{\epsilon^2-\Delta^2}{\epsilon_F^2}}}.
	\label{eq:qe_qh_appendix}
\end{equation}
The BCS coherence factors are
\begin{align}
	u_0(\epsilon)
	&=
	\sqrt{
		\frac{1}{2}
		\left(
		1+\sqrt{\frac{\epsilon^2-\Delta^2}{\epsilon^2}}
		\right)}
	=
	\sqrt{\frac{\Delta}{2\epsilon}}e^{h(\epsilon)/2},
	\nonumber\\
	v_0(\epsilon)
	&=
	\sqrt{
		\frac{1}{2}
		\left(
		1-\sqrt{\frac{\epsilon^2-\Delta^2}{\epsilon^2}}
		\right)}
	=
	\sqrt{\frac{\Delta}{2\epsilon}}e^{-h(\epsilon)/2},
	\label{eq:u_v_appendix}
\end{align}
with
\begin{equation}
	h(\epsilon)
	=
	\begin{cases}
		\operatorname{arccosh}\left(\epsilon/\Delta\right),
		& \epsilon>\Delta,\\
		i\arccos\left(\epsilon/\Delta\right),
		& \epsilon<\Delta.
	\end{cases}
	\label{eq:h_appendix}
\end{equation}
We also define
\begin{equation}
	\Gamma=u_0^2-v_0^2,
	\label{eq:Gamma_appendix}
\end{equation}
and
\begin{equation}
	\begin{pmatrix}
		\Xi_1 & \Xi_2\\
		\Xi_3 & \Xi_4
	\end{pmatrix}
	=
	\begin{pmatrix}
		q_{\tilde h} & q_{\tilde e}\\
		q_{\tilde h} & q_{\tilde e}
	\end{pmatrix}u_0^2
	+
	\begin{pmatrix}
		q_{\tilde e} & q_{\tilde h}\\
		-q_{\tilde e} & -q_{\tilde h}
	\end{pmatrix}v_0^2.
	\label{eq:Xi_appendix}
\end{equation}
For compactness introduce the common denominator
\begin{equation}
	D=
	\Gamma q_{\tilde e}q_{\tilde h}
	+
	\Gamma k_e k_h
	+
	\Xi_1 k_e
	+
	\Xi_2 k_h.
	\label{eq:D_NS_appendix}
\end{equation}
The transparent-interface scattering amplitudes are then~\cite{Blasi2023Hybrid}
\begingroup
\footnotesize
\begin{subequations}
	\label{eq:NS_coefficients_appendix}
	\begin{align}
		r_{ee}
		&=
		\frac{e^{2ixk_e}}{D}
		\left(
		-\Gamma q_{\tilde e}q_{\tilde h}
		+\Gamma k_e k_h
		+\Xi_1 k_e-\Xi_2 k_h
		\right),
		\nonumber\\
		r_{eh}
		&=
		\frac{2u_0v_0\sqrt{k_e k_h}(q_{\tilde e}+q_{\tilde h})}{D}
		e^{i(xk_e-xk_h+\phi)},
		\nonumber\\
		r_{he}
		&=
		\frac{2u_0v_0\sqrt{k_e k_h}(q_{\tilde e}+q_{\tilde h})}{D}
		e^{i(xk_e-xk_h-\phi)},
		\nonumber\\
		r_{hh}
		&=
		\frac{e^{-2ixk_h}}{D}
		\left(
		-\Gamma q_{\tilde e}q_{\tilde h}
		+\Gamma k_e k_h
		-\Xi_1 k_e+\Xi_2 k_h
		\right),
		\nonumber\\
		t_{\tilde e e}
		&=
		\frac{2u_0(q_{\tilde h}+k_h)\sqrt{\Gamma k_e q_{\tilde e}}}{D}
		e^{\frac{i}{2}(-2xq_{\tilde e}+2xk_e-\phi)},
		\nonumber\\
		t_{\tilde e h}
		&=
		-\frac{2v_0(k_e-q_{\tilde h})\sqrt{\Gamma k_h q_{\tilde e}}}{D}
		e^{-\frac{i}{2}(2xq_{\tilde e}+2xk_h-\phi)},
		\nonumber\\
		t_{\tilde h e}
		&=
		-\frac{2v_0(k_h-q_{\tilde e})\sqrt{\Gamma k_e q_{\tilde h}}}{D}
		e^{\frac{i}{2}(2xq_{\tilde h}+2xk_e-\phi)},
		\nonumber\\
		t_{\tilde h h}
		&=
		\frac{2u_0(q_{\tilde e}+k_e)\sqrt{\Gamma k_h q_{\tilde h}}}{D}
		e^{ixq_{\tilde h}-ixk_h+i\phi/2},
		\label{eq:NS_coefficients_normal_in_appendix}\\
		r_{\tilde e\tilde e}
		&=
		\frac{e^{-2ixq_{\tilde e}}}{D}
		\left(
		\Gamma q_{\tilde e}q_{\tilde h}
		-\Gamma k_e k_h
		-\Xi_3 k_e+\Xi_4 k_h
		\right),
		\nonumber\\
		r_{\tilde e\tilde h}
		&=
		-\frac{2u_0v_0(k_e+k_h)\sqrt{q_{\tilde e}q_{\tilde h}}}{D}
		e^{-ix(q_{\tilde e}-q_{\tilde h})},
		\nonumber\\
		r_{\tilde h\tilde e}
		&=
		-\frac{2u_0v_0(k_e+k_h)\sqrt{q_{\tilde e}q_{\tilde h}}}{D}
		e^{-ix(q_{\tilde e}-q_{\tilde h})},
		\nonumber\\
		r_{\tilde h\tilde h}
		&=
		\frac{e^{2ixq_{\tilde h}}}{D}
		\left(
		\Gamma q_{\tilde e}q_{\tilde h}
		-\Gamma k_e k_h
		+\Xi_3 k_e-\Xi_4 k_h
		\right),
		\nonumber\\
		t_{e\tilde e}
		&=
		\frac{2u_0(q_{\tilde h}+k_h)\sqrt{\Gamma k_e q_{\tilde e}}}{D}
		e^{\frac{i}{2}(-2xq_{\tilde e}+2xk_e+\phi)},
		\nonumber\\
		t_{e\tilde h}
		&=
		-\frac{2v_0(k_h-q_{\tilde e})\sqrt{\Gamma k_e q_{\tilde h}}}{D}
		e^{\frac{i}{2}(2xq_{\tilde h}+2xk_e+\phi)},
		\nonumber\\
		t_{h\tilde e}
		&=
		-\frac{2v_0(k_e-q_{\tilde h})\sqrt{\Gamma k_h q_{\tilde e}}}{D}
		e^{-\frac{i}{2}(2xq_{\tilde e}+2xk_h+\phi)},
		\nonumber\\
		t_{h\tilde h}
		&=
		\frac{2u_0(q_{\tilde e}+k_e)\sqrt{\Gamma k_h q_{\tilde h}}}{D}
		e^{-\frac{i}{2}(-2xq_{\tilde h}+2xk_h+\phi)}.
		\label{eq:NS_coefficients_super_in_appendix}
	\end{align}
\end{subequations}
\endgroup
These expressions reduce to the usual Andreev reflection amplitudes only after
taking the Andreev approximation, \(\epsilon,\Delta\ll\epsilon_F\), and
restricting to the subgap transparent-interface regime. In
Fig.~\ref{fig:NS} we keep the full finite-gap form above.

\section{Lorentzian and boxcar transmission}

Here we show the TRB for static transport setups other than quantum point contacts. 

\begin{figure}[b]
	\centering
	\includegraphics[width=\linewidth]{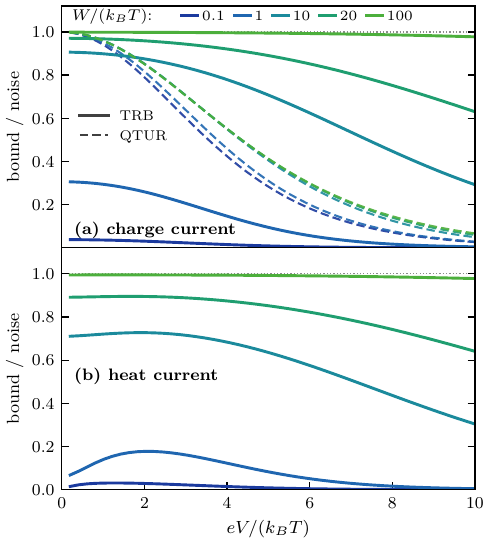}
	\caption{Lorentzian transmission. 
		Solid lines show the TRBs and dashed lines show QTUR bounds.
		Parameters:  $\mu_L=\bar{\mu}+eV/2$, $\mu_R=\bar{\mu}-eV/2$, $T_L=T_R=T$, and
		\(\mathcal{T}(E)=1/[1+\{2(E-E_0)/W\}^2]\), with \((E_0-\bar{\mu})/(k_BT)=0.1\).} 
	\label{fig:SM_Lorentzian}
\end{figure}

Figure \ref{fig:SM_Lorentzian} shows the result when the two terminals are connected by a Lorentzian transmission function. 
In panel (a), charge current measured in  terminal $R$ is considered. 
The TRB (solid) approaches saturation as \(W/(k_BT)\) increases at fixed affinity and dimensionless detuning.
The QTUR bound (dashed) also becomes tighter as the temperature is lowered. However, it does not saturate in the low temperature limit, except in the equilibrium limit.
Panel (b) shows the result for the heat current measured in terminal $R$. 
Here, only the TRB is shown as QTUR is not guaranteed for heat currents. The TRB shows behavior similar to that of temperature dependence as in the charge current.

\begin{figure}[t]
	\centering
	\includegraphics[width=\linewidth]{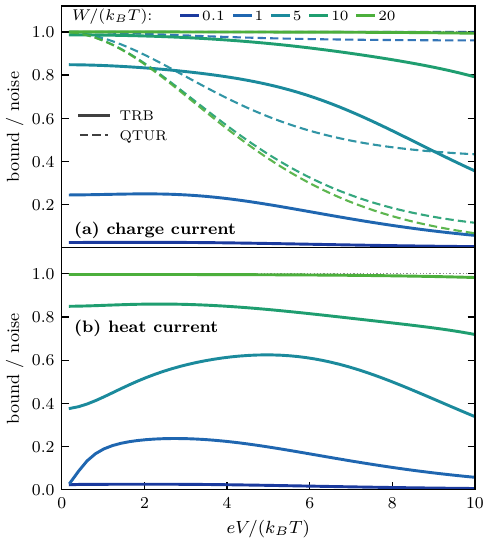}
	\caption{Boxcar transmission. 
		Solid lines show the TRBs and dashed lines show QTUR bounds.
		Parameters:  $\mu_L=\bar{\mu}+eV/2$, $\mu_R=\bar{\mu}-eV/2$, $T_L=T_R=T$, and $\mathcal{T}(E) = 1$ for \(|E-\bar{\mu}|<W/2\) and $0$ otherwise. }
	\label{fig:SM_boxcar}
\end{figure}

Figure \ref{fig:SM_boxcar} shows the result for boxcar transmission function. 
The TRB(solid) shows the similar behavior as in the Lorentzian transmission.
In contrast, the QTUR bound (dashed) shows opposite dependence with respect to temperature: it saturates for high temperatures, i.e., when $k_B T \gg W  $, and becomes less tight when the temperature decreases.

The off-diagonal noise vanishes for the ideal boxcar at all temperatures and becomes negligible for the Lorentzian at low temperature at fixed affinity.
For other transmission profiles, this fraction can remain
finite, keeping the TRB loose.
For isothermal reservoirs with fixed finite affinity
$a=eV/(k_BT)$ and a transmission function smooth near
$\bar{\mu}$, the TRB for the charge current in the low-temperature limit becomes
\begin{equation}
	S\ge B_\textrm{TRB}
	= 2  k_B T \frac{e^2}{h} \mathcal{T}^2(\bar{\mu}) +\mathcal{O}(T^2) ,
\end{equation}
and the bound-to-noise ratio approaches 
\begin{equation}
	\lim_{T\to 0}\frac{B_\textrm{TRB}}{S} = \frac{\mathcal{T}(\bar{\mu})}{ \mathcal{T}(\bar{\mu}) +\{1-\mathcal{T}(\bar{\mu})\} \frac{a}{2} \, \text{coth}\left(\frac{a}{2} \right) } .
\end{equation}
Hence, the TRB saturates in this limit only for unit  transmission. 
At \(a=0\) the factor \((a/2)\coth(a/2)\) has the continuous limit \(1\).

\end{document}